\documentclass[                  
aps,%
12pt,%
final,%
notitlepage,%
oneside,%
onecolumn,%
nobibnotes,%
nofootinbib,%
superscriptaddress,%
noshowpacs,%
centertags%
]{revtex4}

\newcommand{\ab}{Astrophys. Bull. }
\newcommand{\arep}{Astron. Rep. }
\newcommand{\alet}{Astron. Let. }

\begin{document}

\title{The detection of heavy metals  in the circumstellar envelopes \newline of post--AGB stars}
\author{V.G. Klochkova} 
\affiliation{Special Astrophysical Observatory, Nizhnij Arkhyz, 369167 Russia}
          
\date{\today}

\begin{abstract} A new type of peculiarity --- a splitting or asymmetry of strong absorption lines, is found in the optical 
spectra of selected post--AGB stars with  C--rich circumstellar envelopes. The effect is maximal in 
Ba\,II lines whose profile  is split into two--three components. 
The particular components of the split absorption lines are shown to be formed in a structured circumstellar 
envelope, suggesting an efficient dredge--up of the heavy  metals produced during the preceding evolution 
of this star into the envelope. 
We suspect that the splitting (or asymmetry) of the profiles of strongest absorptions with low excitation 
potential of the low level can be associated with the kinematic and chemical properties of  the circumstellar 
environment and with type of its morphology.
\end{abstract}

\maketitle

\section{Introduction}

Now generally accepted that about half of abundances of chemical elements heavier than iron are produced 
by slow neutron  captures (s--process) in the deep layers of asymptotic giant branch (AGB) stars. 
Mixing brings freshly synthesized heavy elements to the stellar surface (the third  dredge-up process) 
and then stellar wind carries them away in the stellar enviroments. This is a simplified description of 
the processes, the sequence of which enriches both circumstellar and interstellar medium with heavy metals. 
For more detail see, e.g., \cite{Herwig,Kappeler} and references there. AGB stars are therefore 
the principal suppliers of heavy metals and important suppliers of carbon and nitrogen to the 
interstellar medium, thereby participating in the  chemical evolution of galaxies. In this article, 
we shall focus on the first observational evidences on the presence of circumstellar spectral features 
of heavy metals in the  optical spectra of nearest  descendants of AGB-stars---protoplanetary nebulae (PPNe).

PPNe  are objects at the  post--asymptotic branch (post--AGB) stage  of evolution.  
These descendants of AGB stars are  low--mass cores with typical masses  of 0.6\,M$_{\odot}$ surrounded 
by an extended and often structured  gaseous--dusty envelope, which formed as a result of substantial mass 
loss by the star during the preceding evolutionary stages. 
Central stars are usually surrounded by envelopes in the form of extended haloes,  arcs,  lobes, and tori. 
Some stars exhibit  various combinations of the above features as well as bipolar and quadrupolar nebulae 
with dust bars. Examples of the latter two types include the Egg nebula\,=\,RAFGL~2688 and  {\it IRAS\/}\,19475+3119 
nebulae, for which high-resolution images  were taken by the Hubble Space Telescope~\cite{Siodmiak}. 
All PPNe and about  80\% of planetary nebulae are asymmetric~\cite{Lagadec}. Much remains to be understood 
about condensation of dust particles and formation of the dust fraction in AGB star envelopes 
(see~\cite{Bieging} and references therein).

A circumstellar gas and dust envelope shows up in the form of peculiarities in the IR, radio, and optical 
spectra of post-AGB supergiants. The optical spectra of PPNe differ from those of classical massive supergiants 
by the presence of molecular bands superimposed onto the spectrum of an F--G supergiant and by the anomalous 
behavior of the profiles of selected spectral features. These may include complex emission and absorption profiles 
of H\,I, Na\,I, and He\,I lines,  and metal emission features (see for detail~\cite{Envelope} and references 
therein). 
Furthermore, all these peculiarities are variable. In whole we see that the known types of spectral features 
in the optical spectra of post--AGB stars are: 
1) low-- or moderate--intensity  symmetric metal absorptions without apparent distortions; 
2) complex profiles of neutral hydrogen lines,  which vary with time and include absorption and emission components; 
3) absorption or emission bands of mostly carbon  containing molecules; 4) envelope components of the  Na\,I 
and K\,I resonance lines, and also 5) narrow permitted or forbidden metal emission lines that form in envelopes. 
The presence of type 2--5 features is the key difference  of the spectra of PPNe from those of massive supergiants.

Here we analyze the manifestations of circumstellar envelopes in the optical spectra of  PPNe focusing  on the  
homogeneous subsample of stars whose atmospheres, according to previous studies, underwent evolutionary 
variations of the  chemical composition. Section~2 briefly describes the employed observational 
data and lists the studied  stars and their basic parameters. In Section\,3 
we analyze the available data on the peculiarities  of the profiles of metal lines, found in high-resolution spectra, 
as well as data on the presence of molecular bands and outflow velocities for objects with different envelope structures. 
In Sections~4  and~5 we discuss the obtained results  and summarize the main conclusions.

\section{OBSERVATIONAL DATA}

Over the past two decades  more than 40 post-AGB  candidates --  supergiants with IR excesses and several related 
luminous stars with  unclear evolutionary status have been spectroscopically monitored with the 6--m telescope 
of the Special Astrophysical Observatory. As a result, a collection of high-quality spectra 
has been acquired with the primary  purpose of searching for anomalies of stellar chemical composition due 
to the nucleosynthesis of chemical elements in the interiors of low- and intermediate-mass stars  and the 
subsequent dredge-up of the synthesis products to the surface layers of stellar atmospheres. These observational
data are also used to search for peculiarities in the PPNe spectra, to analyze the velocity fields in the atmospheres  
and envelopes of these stars with mass loss, and to search for the likely long-term spectral and radial velocity pattern  
variations.

Here we use the data acquired in the Nasmyth focus with the NES~\cite{nes} echelle spectrograph. 
The NES spectrograph, equipped with a 2048$\times$2048 CCD and an image slicer, 
produces a spectroscopic resolution of R$\approx60\,000$. Since 2011 the NES spectrograph has been equipped 
with a 2048$\times$4096  CCD which made it possible to significantly extend the wavelength coverage.  
The spectra of the faintest program objects (the optical component of the IR sources {\it IRAS}\,04296+3429 and 20000+3239) 
were acquired with the PFES  echelle spectrograph mounted in the 
primary focus of the 6-m telescope~\cite{pfes}.  This spectrograph,  equipped with a 1k$\times$1k CCD,  
produces a spectroscopic resolution of R$\approx$15\,000. We described  the details of spectrophotometric 
and position measurements of the spectra in our earlier papers, the corresponding references can be found  
in the original papers listed in the paper by~\cite{Envelope}.

\begin{table}
\caption{Basic data for C--rich circumstellar envelopes of post-AGB stars. Details concerning  the C$_2$ bands 
         in the third column see in the text. The last column  gives the expansion velocity of the envelope  
         as determined  from the position of  Ba\,II circumstellar components.}
\medskip
\begin{tabular}{> \small c|> \small l|> \small l| > \small l|> \small l|> \small l}
\hline
 Object  &   Morphology           & Type of  & \multicolumn{3}{c}{\small $V_{\rm exp}$, km\,s$^{-1}$} \\
\cline{4-6}
         & of the                 & the C$_2$&           &         &   \\[-10pt]
         &    envelope$^a$        &  bands   &  CO       & C$_2$   &  BaII   \\
\hline
04296+3429& bipolar\,+            & abs      & 10.8$^b$  & 7.7$^g$  & \\  [-10pt]
          & halo\,+  bar          & emis     &           & 12$^h$   & \\ 
\hline
07134+1005& elongated             & abs      & 10.2$^b$  & 8.3$^g$  & \\   [-10pt]
          &   halo                & abs      &           & 11$^i$   & \\ 
\hline         
08005$-$2356& bipolar             & uncertain& 100:$^c$  & 43.7$^g$ & \\  [-10pt]
          &                       & abs      &           & 42$j$    & \\ 
\hline
19500$-$1709& bipolar             & no       &17.2, 29.5$^b$&       &20 \& 30$^k$ \\ [-10pt]
          &                       &          &10, 30--40$^d$&       & \\     
\hline          
20000+3239& elongated             & abs      & 12.0$^e$  & 12.8$^g$ & \\ [-10pt]
          & halo                  & abs      &           &  11.1$^l$& \\ 
\hline        
RAFGL\,2688 &multipolar\,+        & abs      &17.9, 19.7$^f$& 17.3$^g$&\\ [-10pt]
          & halo\,+\,arcs         & emis     &           & 60$^m$   & \\ 
\hline          
22223+4327& halo\,+               & abs.     & 14--15$^f$& 15.0$^g$ & \\ [-10pt]
          &  small lobes          & emis     &           & 15.2$^n$ & \\ 
\hline          
22272+5435&elongated\,+           & abs      &9.1--9.2$^b$& 9.1$^g$ & \\ [-10pt]
          & halo\,+\,arcs         & abs      &            &  10.8$^o$&10$^o$ \\ 
\hline          
23304+6147& quadrupole\,+         & abs      &9.2--10.3$^b$& 13.9$^g$& \\ [-10pt]
          & halo\,+\,arcs         & emis    &             & 15.5$^p$&15.1$^p$\\ 
\hline 
\multicolumn{6}{l}{\small \it Notes:  a -- morphology type of envelopes is taken from papers by \cite{Ueta2000,Sahai,Siodmiak,Lagadec};} \\  [-10pt]        
\multicolumn{6}{l}{\small \it b -- \cite{Hrivnak},  c -- \cite{Hu}, d -- \cite{Bujar}, e -- \cite{Omont},  f -- \cite{Loup}, g -- \cite{Bakk97},}\\ [-10pt]
\multicolumn{6}{l}{\small \it h --  \cite{04296},   i --\cite{atlas}, j -- \cite{08005}, k -- \cite{19500}, l -- \cite{20000}, m -- \cite{Egg1},} \\ [-10pt]
\multicolumn{6}{l}{\small \it n -- \cite{22223},  o -- \cite{V354Lac}, p -- \cite{23304b}.}  \\
\end{tabular}
\label{PPN}
\end{table}

\section{Main peculiarities in the optical spectra of post--AGB stars }

Our comprehensive study of the program stars allowed us to determine (or refine) their evolutionary status. 
One of the results of our analysis is that the studied sample of luminous stars with IR excesses is not
homogeneous~\cite{VAK}. In this paper we consider  the peculiarities of the optical spectra of  
post--AGB stars paying special attention to the subsample of objects listed in Table\,\ref{PPN}. 
It contains  {\it IRAS} objects  with central stars whose atmospheres are overabundant in carbon and 
heavy metals. Their circumstellar envelopes have a complex morphology and are usually rich in carbon, 
as evidenced by the presence of C$_2$,  C$_3$, CN, CO, etc. molecular bands in their IR, radio, and 
optical spectra. Presence and type of  C$_2$ bands for stars in Table\,\ref{PPN} are published by 
\cite{Bakk97,19500,08005,20000};  \cite{04296,Egg1,atlas,V354Lac,22223,23304b}.

Furthermore, the objects from Table\,\ref{PPN} are among those few PPNe whose IR--spectra exhibit the so far 
unidentified emission band at 21\,$\mu$m~\cite{Kwok-21,Hrivnak2009}. Despite an extensive search for appropriate 
chemical agents, so far no conclusive identification has been proposed for this rarely observed feature. 
However, its very presence in the spectra of PPNe with carbon enriched envelopes suggests that this emission 
may be due to the presence of a complex  carbon-containing molecule in the envelope (see~\cite{Hrivnak2009,Li} 
for details  and references).

Overabundances of carbon and heavy-metal [s$/$Fe] overabundance (or lack thereof) in the atmosphere of the 
central star were published earlier by \cite{04296,07134,08005,23304} for  {\it IRAS}\,04296+3429, 
07134+1005, 08005$-$2356, and 23304+6147 respectively;  \cite{20000} for  {\it IRAS}\,20000+3239;  \cite{Egg2} and 
\cite{Ishigaki} for RAFGL~2688; \cite{22223} for V448~Lac and \cite{V354Lac} for V354~Lac. Chemical abundances 
for five stars from Table\,\ref{PPN} were also published by~\cite{Reyn}. As follows from Table\,\ref{Split}, 
all stars with excess  carbon and heavy metals are objects with moderate iron deficiency and may belong to 
the thick disk of the Galaxy.

The spectra of protoplanetary nebulae with F--K-type supergiant as central stars that have carbon-enriched atmospheres 
show features of carbon-containing molecules  C$_2$, C$_3$,  CN, and CH$^+$.  Position  measurements of molecular 
features in the spectra indicate that they form in expanding circumstellar envelopes. Authors~\cite{Bakk97} and 
\cite{04296, atlas, 08005, 20000, Egg1, 22223, V354Lac, 23304b} used high-resolution optical spectra to analyze  
molecular bands for several post--AGB stars including some objects from Table\,\ref{PPN}.  

It appears that the emission in the Swan bands or Na\,I D--lines is observed in the spectra of PPNe with bright 
and conspicuously asymmetric circumstellar nebulae. The results of spectroscopic observations of several PPNe 
confirm this hypothesis. An analysis of the spectra taken with the 6--m telescope revealed  C$_2$ Swan 
emission bands of different intensities (relative to the continuum) in the spectra of the central stars of the
following sources:  {\it IRAS}~04296+3429~\cite{04296}, 08005$-$2356~\cite{08005},
$RAFGL$~2688~\cite{Egg1},  {\it IRAS}~22223+4327~\cite{22223}, and   {\it IRAS}~23304+6147~\cite{23304,23304b}. 
According to HST images~\cite{Ueta2000,Siodmiak}, these objects have structured (and often bipolar) envelopes. 

In addition to a sample of related C--rich objects with the above features, Table\,\ref{PPN} also includes the infrared
source   {\it IRAS}~08005$-$2356. This object   has so far been  poorly studied, and no data  are available either 
on the peculiarities of the chemical composition of its atmosphere or on the presence  of the 21-$\mu$m band. 
However,  {\it IRAS}~08005$-$2356  can be viewed as related to  the objects of this sample 
because its optical spectrum  exhibits C$_2$~Swan bands, the hydrogen and  metal lines in its spectrum have 
emission-absorption profiles~\cite{08005}, and the circumstellar  envelope is observed in CO emission~\cite{Hu}. 

Note that our subsample of PPNe,  thoroughly studied by high-resolution spectra (Table\,\ref{PPN}), practically 
coincides with  the list of C--rich and 21\,$\mu$m  protoplanetary nebulae the photometric and spectral properties 
of which were extensively studied in papers~\cite{Hrivnak2009,Hrivnak2010}. Of fundamental importance to us is the 
conclusion in~\cite{Hrivnak2011}  about the rare occurrence of binaries among post--AGB stars of the type considered, 
which is based on the long-term  investigation of the velocity field in PPN atmospheres conducted by the above authors.
Thus, the results of \cite{Hrivnak2009,Hrivnak2010,Hrivnak2011} provide further evidence for the homogeneity of 
the PPNe subsample considered here.

\subsection{Peculiar metal absorptions in the optical spectra of selected post--AGB stars}

\begin{figure*}
\includegraphics[angle=-90,width=114mm,bb=30 80 550 790,clip]{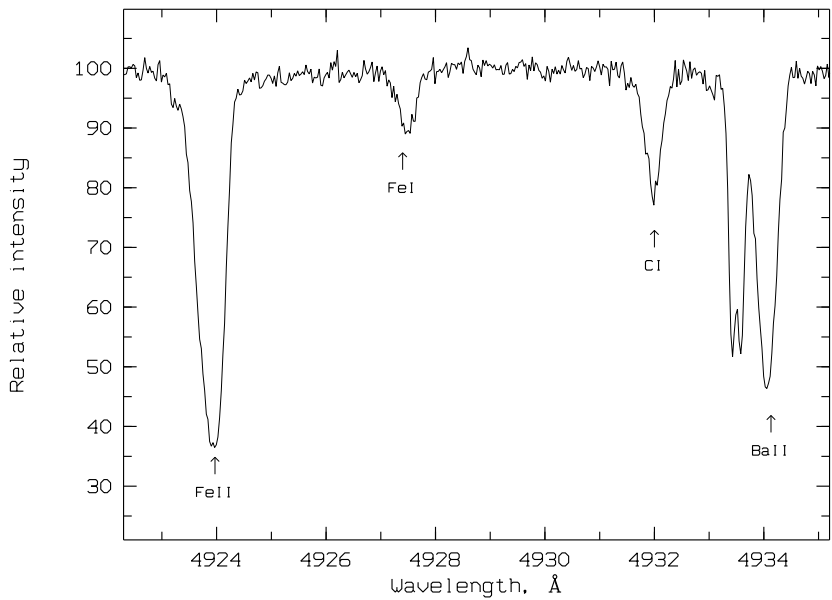} % {V5112Sgr-Ba4934.ps} 
\caption{Fragment of the spectrum for V5112~Sgr containing split Ba\,II\,4934\,\AA{} line  in the July 7, 2001 spectrum.  
     The identification of main absorptions is indicated.}
\label{V5112Sgr-Ba4934fr}
\end{figure*}

The systematic monitoring of PPN candidates, which we performed with a high spectral resolution, has  
released a new result---an unknown earlier peculiarity: the splitting (or asymmetry) of strongest metal  
absorptions due to distortion by envelope  features.  Now such peculiar profiles  of the strongest absorptions 
were revealed in the spectra of five stars listed in Table\,\ref{Split}: CY~CMi~\cite{atlas}, 
V354~Lac~\cite{V354Lac,V354LacK}, V448~Lac~\cite{22223}, V5112~Sgr~\cite{19500}  and  CGCS~6918 
({\it IRAS}~23304+6147)~\cite{23304b}. Let us illustrate this effect using the example of the spectrum of the 
high-latitude supergiant V5112~Sgr,  where  it is most  pronounced. Figure\,\ref{V5112Sgr-Ba4934fr} shows
a fragment of the spectrum of   V5112~Sgr taken on July~7, 2001 with a split Ba\,II~4934~\AA{} line.  
At the same time,  the profiles of strong absorptions of iron-group metals in the spectrum of this star 
are neither asymmetric nor split, as is immediately apparent from the same Fig.\,\ref{V5112Sgr-Ba4934fr}, 
where we see a very strong but unsplit Fe\,II~4924~\AA{} line.

 %{V5112Sgr-Ba4934.ps} 

\begin{figure}
\includegraphics[width=74mm,bb=40 70 550 790,clip]{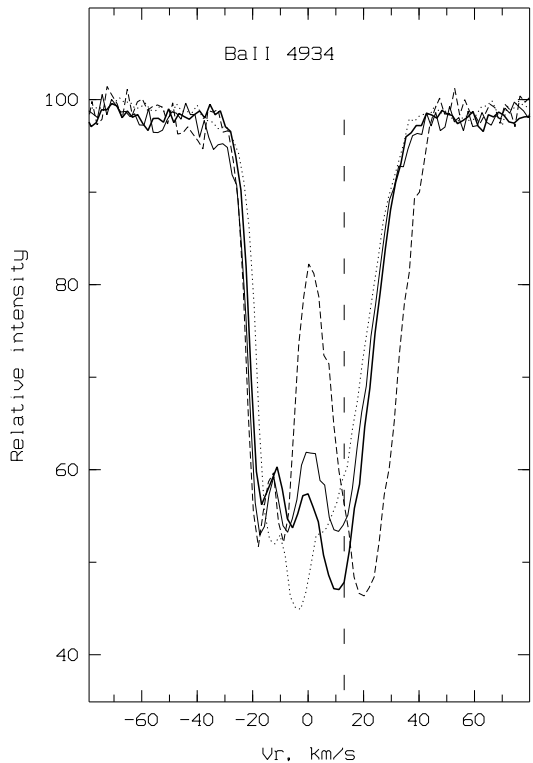}  
\caption{Variability of the  Ba\,II~4934~\AA{} line profile in the spectra of V5112~Sgr taken in different years: 
         August 2, 2012 (the thin solid line); June 13, 2011 (the solid bold line); August 14, 2006 (the dotted line); 
         July 7, 2001 (the dashed line)~\cite{19500}.} 
\label{V5112Sgr-Ba4934}
\end{figure}

Besides, a comparison of the line profiles in the spectra of V5112~Sgr taken during different nights 
reveals substantial variability of the profile shapes and of the positions of the components of the split lines. 
To illustrate the variability effect, we show in Fig.\,\ref{V5112Sgr-Ba4934} the  Ba\,II~4934~\AA{} line profile, 
which is most asymmetric and most variable. The different widths of the components are  immediately apparent in both 
Figs.\,\ref{V5112Sgr-Ba4934fr} and \ref{V5112Sgr-Ba4934}: the red component is about twice  broader than the blue 
components, which are offset substantially relative to  the systemic velocity. This difference between the component 
widths indicates that the red and blue components form under different physical conditions. It follows also from  
Fig.\,\ref{V5112Sgr-Ba4934fr}  that the position of the photospheric (red) component of the complex profile is variable,
whereas the blue components, which, as shown in~\cite{19500},  form in the envelope, are stable. 

 %{V5112-prof}

 %{IRAS22223-prof}
\begin{figure}
\includegraphics[width=74mm,bb=30 70 550 780,clip]{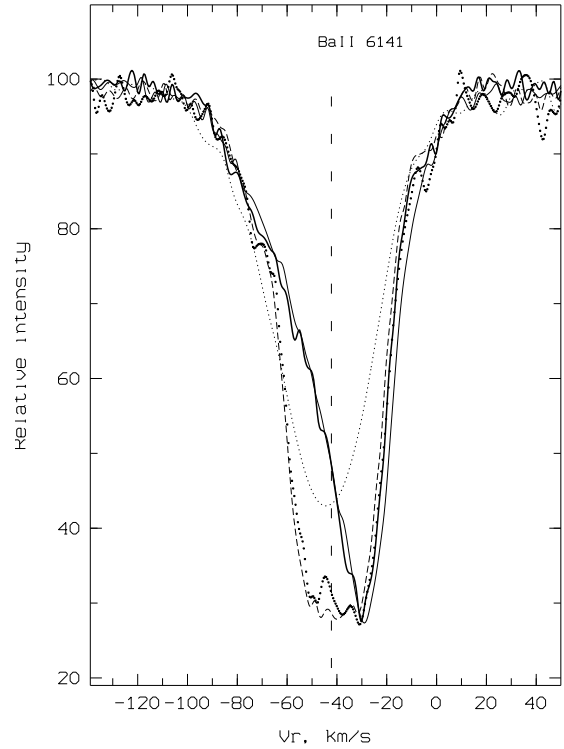}   
\caption{Profile variations of the BaII\,6141\,\AA{} line in the spectra of V448~Lac at various epochs: 
         JD\,2454760.17 (bold), JD\,2454721.15 (thin), JD\,2453694.36 (dashed), and JD\,2452131.53 (dotted)} 
\label{V448Lac-Ba6141}
\end{figure}

%{IRAS23304-prof} 
\begin{figure}
\includegraphics[width=64mm,bb=30 70 550 780,clip]{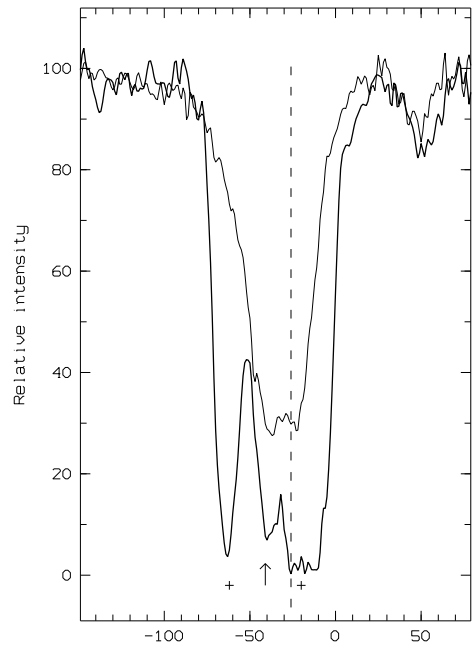}   
\caption{Profiles of the split lines in the October 12, 2013 spectrum of CGCS~6918: D$_2$\,NaI -- the bold line 
           and Ba\,II\,6141 -- the the thin line}
\label{Profiles-23304}
\end{figure}

\begin{table}
\caption{Main  atmospheric parameters of post--AGB stars with peculiar profiles of strongest absorptions: effective temperature, Teff,
         metallicity, [Fe/H]$_{\odot}$, and the mean excess of heavy metals synthesized in s--process, [s/Fe]$_{\odot}$}
\medskip
\begin{tabular}{> \small r > \small  c > \small c > \small c> \small  c > \small c}
\hline
 Star      & {\it IRAS}  & Teff, {\it K} & [Fe/H]$_{\odot}$ & [s/Fe]$_{\odot}$ & Ref \\
\hline
CY~CMi     & 07134+1005  & 7000    &$-1.0$   & +1.4  & a \\   [-10pt]        
V5112~Sgr  & 19500$-$1709& 8000    &$-0.6$   & +1.1  & b\\   [-10pt]             
V448~Lac   & 22223+4327  & 6500    &$-0.3$   & +0.9  & b \\    [-10pt]             
V354~Lac   & 22272+5435  & 5650    &$-0.8$   & +1.2  & c \\    [-10pt]        
CGCS~6918  & 23304+6147  & 5900    &$-0.6$   & +1.2  & d  \\  
\hline 
\multicolumn{6}{l}{\small \it  a -- \cite{07134}, b -- \cite{Reyn}, c -- \cite{V354Lac}, d -- \cite{23304}.} \\  
\end{tabular}
\label{Split}
\end{table}

The semiregular variable V354~Lac is the closest analog to V5112~Sgr among the objects listed in Tables\,\ref{PPN} 
and \ref{Split} in terms of the structured envelope and chemical abundances. The spectroscopic monitoring of 
this star~\cite{V354Lac}  carried out at the SAO 6--m telescope with a resolution of $R=60\,000$ also revealed 
the splitting of the strongest  absorptions with the low-level excitation potential of  $\chi_{\rm low}\le 1$~eV. 
An analysis of the kinematic pattern showed that the blue component of the split line forms in the powerful gas 
and dust envelope of V354~Lac.  This splitting shows up most conspicuously in the profile of the strong 
Ba\,II~6141~\AA{} line. The shift of the blue  component of the Ba\,II line coincides with that of the circumstellar 
component of Na\,I D lines, which forms in the same  layers as the circumstellar C$_2$ Swan bands. This coincidence 
indicates that the complex profile of the Ba\,II~6141~\AA{}  line contains, in addition  to the atmospheric component, 
a component that forms in the circumstellar envelope. 
Such splitting (or the profile asymmetry   due to the more shallow slope of the blue wing) is also observed 
for other Ba\,II ($\lambda\lambda$\,4554, 5853, 6496~\AA{})  lines as well as for the strong Y\,II~5402~\AA{}, 
La\,II~6390~\AA{}, and Nd\,II~5234, 5293~\AA{} lines. The lines of these  ions in the spectrum of V354\,Lac are 
enhanced to the extent that  their intensities are comparable to those of neutral-hydrogen  lines. For example, 
the equivalent width of the Ba\,II~6141~\AA{}  line reaches $W_{\lambda}\approx1$~\AA{}, and that of H$\beta$ 
reaches $W_{\lambda}\approx2.5$~\AA{}.

As follows from Figs.\,\ref{V448Lac-Ba6141} and \ref{Profiles-23304}, in the spectra of two stars---V448~Lac and 
CGCS~6918 -- a central star of  {\it IRAS}\,23304,  the Ba\,II~6141~\AA{}  line profile is or asymmetric or split for
different moments of observations. Note that the variations are displayed only by the short-wavelength profile 
wings and the position  of the core, whereas the positions and intensities of the long-wavelength wings of these 
strong Ba\,II, La\,II, and  Y\,II absorption lines do not  vary in time.

Asymmetric and variable profiles of strongest lines  (Y\,II, Ba\,II, Fe\,II and other strong absorption lines) 
formed in the expanding stellar atmosphere were also found in~\cite{56126} in the spectra of CY~CMi.   
However, as follows from Fig.\,5 in the paper by these authors, in the spectra of  CY~CMi profiles are just 
only asymmetric, but not split. Thus, we see a difference between the peculiarity types of the profiles of 
three stars---V5112~Sgr, V354~Lac, and  CGCS~6918 with split profiles of the strongest 
absorptions of selected elements, and two stars---CY~CMi and V448~Lac with asymmetric but unsplit profiles. 
Such a difference permit us to suggest that the morphology of the circumstellar envelope may  be the main 
factor that causes the peculiarity and variability of the profiles of the strongest lines. As is evident  
from Table\,\ref{PPN}, the two stars V5112~Sgr, V354~Lac and CGCS~6918 with split absorptions have bipolar 
(or quadrupolar) envelopes, whereas the absorptions are unsplit in the spectra of  CY~CMi  and V448~Lac 
the envelope of these stars have a less structured environment. This hypothesis is further corroborated 
by the three-component structure of the  strong absorption profiles observed in the spectrum of  V5112~Sgr,  
where CO--observations show both the slow (V$_{\rm exp}$\,=\,10\,km\,s$^{-1}$) and the fast 
($30$--$40$\,km\,s$^{-1}$)  expansion~\cite{Bujar}. As follows from data in Table\,\ref{PPN}, 
the profiles  of the split lines include a photospheric component and  two envelope components, one of which, 
like in the case of  the CO--profile, arises in the envelope that formed at  the AGB--stage  and expands at 
a velocity of  V$_{\rm exp}(2)\approx 20$~km\,s$^{-1}$, and the other one arises in the 
envelope that moves at a velocity of  V$_{\rm exp}(1)\approx 30$~km\,s$^{-1}$ and formed later.

Worth mentioning here is another interesting phenomenon which was revealed in the spectrum of V5112~Sgr.  
Klochkova~\cite{19500}  showed that the optical spectrum of this high latitude  post--AGB supergiant  contains
weak absorptions  whose positions are indicative of their formation in the circumstellar envelope. The velocities 
V$_r$(DB) were measured  in the 5780--6379~\AA{} wavelength interval from the positions of reliably identified 5780, 
5797, 6196, 6234, and 6379~\AA{} features. The mean V$_r$(DIB) averaged over several spectra and determined with 
an accuracy better than $\pm 0.5$~km\,s$^{-1}$  agrees excellently with the velocity determined from the blue 
component of the Na\,I D--lines. The resulting agreement leads us to conclude that the weak bands found in the 
spectrum of V5112\,Sgr origin in the circumstellar envelope. Kipper~\cite{Kipper2013} independently came to 
similar conclusions for the same object based on the spectra of V5112~Sgr taken with a different instrument. 
This DIBs identification at the present time is a unique result in the task of searching for these spectral 
features in the circumstellar environment.

\section{Discussion of the results} 

Note that the splitting of strong absorptions were observed in the spectra only those post--AGB stars that 
are listed in the Table\,\ref{Split}, in atmospheres which were found significant excess carbon and heavy metals.
The second precondition  for the splitting of the strong absorptions in the spectrum  of these post--AGB stars  
is presence  of  a complex structured envelope.  We can  suggest that the process  of the formation 
of a circumstellar envelope may contribute to the enrichment of this envelope by products of stellar  nucleosynthesis. 
The profiles of the split lines contain a photospheric and one--two  envelope components, one of which,  
like in the case of the  CO--profile, arises in the envelope that formed at the AGB--stage, and the 
other one---in the envelope that formed later.  The circumstellar components of  strong heavy-metal absorptions 
have  been conclusively identified in the spectra of V5112~Sgr~\cite{19500}, V354~Lac~\cite{V354Lac} and 
CGCS~6918~\cite{23304b}. 

The structure of the circumstellar nebulae of  V354\,Lac and CGCS\,6918  may be more complex than it appears 
in the HST observations (see Table\,\ref{PPN}). Polarimetric observations of  V354\,Lac~\cite{Gledhill} 
indicate  the presence of a ring structure embedded in an extended nebula. Nakashima et al.~\cite{Nakashima2012} point out 
that the axes of the optical and infrared images of the nebula are almost perpendicular to each other. 
Based on the kinematic pattern of the nebula as determined from the mapping of the CO--emission, 
authors~\cite{Nakashima2012} concluded that the structure of the nebula includes not only a torus and a spherical 
component but also another element (possibly a jet).

Currently, no consensus has been reached concerning the development of deviations from spherical symmetry 
in PPNe. Authors~\cite{Ueta2000, Siodmiak} analyzed high spatial resolution optical images of a sample 
of PPNe taken by the Hubble Space Telescope and concluded that the optical depth of the circumstellar matter 
is the crucial factor that determines the formation of a particular morphology of stellar envelopes. 
The dense and often spherical envelope that formed during the AGB--stage is believed 
to expand slowly, whereas the rapidly expanding feature is the axisymmetric part of the envelope that 
formed later at the post--AGB stage~\cite{Castro2010}. 
The sequence of these processes results in the development of an optical depth gradient in the direction 
from the equator to the polar axis of the system. The presence of a companion and/or a magnetic field in
the system may also prove to be the physical factor that causes the loss of the spherical symmetry of 
the stellar envelope during the short evolutionary interval between the AGB and post--AGB stages 
(see~\cite{Huggins,Ferreira} and references therein). 
In their recent paper~\cite{Koning} proposed a simple PPN model based on a pair of evacuated cavities 
inside a  dense spherical halo. The above authors demonstrated that all the morphological features observed 
in real bipolar PPNe can be  reproduced by varying the available parameters (mass density inside the cavity, 
its size and orientation) of this model.

So far, the discovery of the heavy-metal enrichment of the circumstellar envelopes of the post--AGB supergiants 
V5112~Sgr, V354~Lac and CGCS~6918  remain the only results. Here it should be recalled that Mauron \& Huggins~\cite{Mauron} 
detected atomic metals (Ca, Fe and upper limits for Al, Ti, Mn and Sr) in the C--rich circumstellar envelope 
of the AGB--star CW~Leo, which is a central star of the IR source  {\it IRC}\,+10216.  

The effect of splitting of metallic lines found requires to continue the high resolution spectroscopy of  
very related  post--AGB stars.  The most promising objects could be  {\it IRAS}~04296+3429 and {\it RAFGL}~2688. 
It follows from Table\,\ref{PPN}  that the totality of properties of their envelopes coincide with 
those of the objects  listed in Table\,\ref{Split}:  they  are C--enriched  and have a very complex structure. 
Besides, the atmospheres of the weak central stars of both  sources are enriched in heavy metals~\cite{04296,Egg1}.  

Evidently, the next step in study of the discovered splitting of the profiles  should be theoretical modeling 
of the spectra envelopes and calculation chemical abundances, which could be consider with regard to selective 
depletion of chemical elements  in a dusty  environment.

\section{Main coclusions}

We used the results of high spectral resolution observations made with the 6--m telescope to analyze 
the peculiarities of the optical spectra of a sample of post--AGB stars with atmospheres enriched in carbon and heavy
s--process metals and with carbon-enriched circumstellar envelopes.

We showed that presence of the peculiarities of the line profiles (the asymmetry and splitting of the profiles of strong absorptions) 
is  associated with the kinematic and chemical properties of the circumstellar envelope and the type of its morphology. 
The splitting of the profiles of the strongest heavy-metal absorptions in the spectra of the  V5112~Sgr, V354~Lac
and CGCS~6918  supergiants  found as a result of our observations suggests that the formation of a structured circumstellar 
envelope is accompanied by the  enrichment of this envelope with the products of stellar nucleosynthesis. 

Attempts to find a definite link between the peculiarities of the optical spectrum and the morphology of the circumstellar 
environment are complicated by the fact that the observed structure of the envelope depends strongly on the inclination 
of the symmetry axis to the line of sight and on the angular resolution of the spectroscopic and direct-imaging instruments.

In the spectrum of the high latitude post--AGB supergiant V5112\,Sgr were revealed weak absorptional DIBs  5780, 
5797, 6196, 6234, and 6379~\AA{}.  Their mean radial velocity V$_r$(DIBs) determined with  an accuracy better than 
$\pm 0.5$~km\,s$^{-1}$  agrees excellently with the velocity determined from the blue circumstellar component of 
the Na\,I D--lines.

\section*{Acknowledgments}

This work was supported by the Russian Foundation for Basic Research (project No.\,14--02--00291\,a). 
This research has made use of the SIMBAD database, operated at CDS, Strasbourg, France, and NASA's Astrophysics Data System.

\end{document}